# Improved interfacial resistance and crystal-structure stability in a low-cobalt P2-type sodium-ion battery cathode material

William Rexhausen[a*], Christian Parsons[a*], Uma Garg[a*], Deyang Qu[b], Prasenjit Guptasarma[a]

[a] *Department of Physics, University of Wisconsin-Milwaukee, 3135 N. Maryland Avenue, Milwaukee, WI, 53211, USA*

[b] *Mechanical Engineering Department, University of Wisconsin-Milwaukee, 3200 N Cramer St, Milwaukee, WI, 53211, USA*

*These authors contributed equally

**Abstract**

We describe $Na_{0.67}Mn_{0.625}Fe_{0.25}Co_{0.125}O_2$ (NMFCO), a P2-type sodium-ion battery cathode. Our composition, with significantly less Co than in an earlier study, shows discharge capacity close to 190 mAhg$^{-1}$ and specific energy density exceeding 500 mWhg$^{-1}$ in the 1.5 to 4.3 V range. The material also shows an improved structural stability over similar materials. Such changes, between the pristine phase ($P6_3/mmc$, $P6_3$ (OP4), or orthorhombic *Cmcm*) and the so-called "Z" phase, are endemic to other P2-type cathodes such as $Na_{0.67}Mn_{0.65}Fe_{0.35}O_2$ (NMFO). We propose two equivalent circuit models of impedance spectroscopy to understand electrochemical processes in our cells with a sodium metal anode. Our equivalent circuit modeling, combined with an analysis of the initial galvanostatic slope, reveals a significant reduction in the polarization and interfacial charge-transfer resistance at the solid electrolyte interface. We reveal that the combined effects of crystal structure stability, lower internal resistance, relatively high specific energy density, and improved battery health make this low-cobalt P2-type cathode composition a very promising candidate for new sodium-ion batteries.

**Highlights**

1. Co-substitution in Na-ion cathode NMFCO suppresses distortions at very low Co level.
2. Over 190 mAhg$^{-1}$ capacity and 500 mWhg$^{-1}$ energy density in 1.5-4.3 V cycling range.
3. EIS-based equivalent circuit modeling shows lower interfacial resistance in NMFCO.
4. Cycling does not greatly degrade State of Health or internal resistance of NMFCO.

**Keywords:** Energy Storage; Na-ion cathode; Structural transition; Rietveld refinement; Layered oxides; Electrochemical Impedance Spectroscopy; Interfacial Impedance; Equivalent Circuit Modeling.

## 1. Introduction

The low cost and relative abundance of sodium resources makes sodium-ion batteries an attractive alternative to lithium-ion batteries in specific applications such as large-scale energy storage [1], [2], [3], [4]. Disruptions to the limited supply chain of Li resulting from, for example, political disturbances or worldwide shutdowns such as with the CoViD19 pandemic of 2020, can critically harm Li-ion battery cathode manufacturing and the energy economy [5]. It is therefore important to explore new compositions for high-performance Na-ion electrode materials which balance high energy density and cyclability with mineral availability and cost.

The study of sodium ion-transition metal oxide cathodes dates to the earliest research on lithium and sodium ion batteries. Fouassier et al. studied $Na_xMO_2$ with M = Mn, Co, Cr in the 1970s, and Delmas et al. studied the stacking structure of such compounds in the 1980s, leading to the *Delmas notation* for stacking patterns which we use in this article [6], [7]. More recently, Wang et al. and Meng et al. have studied the phase stability of sodium cobalt oxide. Bucher et al. have reported a relation between morphology and cycling stability in sodium manganese cobalt oxide [8], [9], [10]. In addition, layered oxides in the family $Na_xMO_2$ (where M is any transition metal, or combination of transition metals) have recently gained ground as cathode materials for Na-ion batteries [11], [12], [13]. In this work, we report the effects of cycling on structure, morphology and impedance of $Na_{0.67}Mn_{0.625}Fe_{0.25}Co_{0.125}O_2$ (NMFCO). A similar composition has been explored by one other group [13]. To alleviate broader concerns about the toxicity and cost of Co, we explore the use of a significantly lower concentration of Co than that used by Liu et al [13]. Given that we are still in the early days of Na-ion battery research, it is important that we explore new phases and stoichiometries beyond the question of commercialization. Such study can provide important clues about relationships between crystal-structure, physical properties, mechanisms, and functions, transferable to future commercial designs of batteries.

We report here our results on cyclability and capacity retention in NMFCO by extending charge cycling to 200 cycles, and in different voltage ranges, 1.5-4.3 V, 1.5-4 V and 2-4 V, to identify the effect of cycling in relatively narrow voltage ranges where the lattice stress is reduced. Previous studies of Na-ion battery cathode stoichiometries have been limited to lower charge cycle numbers, such as 30 cycles. Another highlight of this paper is a detailed study and modeling of Electrochemical Impedance Spectroscopy (EIS), a technique which has not yet been widely reported in Na-ion cathode batteries, but which provides important information on the state of health with respect to factors such as cathode-electrolyte interface and the Solid Electrolyte Interface. Finally, we report discharge capacity of $Na_{0.67}Mn_{0.625}Fe_{0.25}Co_{0.125}O_2$ up to 200 cycles and use Rietveld refinement of X-Ray Diffraction (XRD) data to elucidate structural changes. Specifically, our XRD results show that the "Z" phase does not form in the lower Co stoichiometry reported here, in contrast with Liu et al. who report that the "Z" phase forms above 3.6 V. Liu et al. discuss that the "Z" phase, which shows up in several compositions in this P2-type family could be responsible for capacity-fading in Na-ion cathodes [13]. The implication of this finding to additional theories for capacity fading, and the appearance of voltage plateaus in galvanometric cycling, is further discussed in a later section. We provide Nyquist plots showing the in-phase and out-of-phase impedance, as well as plots showing the dependence of total impedance on frequency. Using an in-house modeling routine, we provide quantitative information about the equivalent internal circuit in our batteries (equivalent circuit modeling) and discuss the role of different cell components related with the electrochemistry of the cell. We then compare this with a new method to evaluate the internal resistance of our cells from galvanostatic plots. Our methods and results consistently reveal that the internal resistance arising from a Solid Electrolyte Interface (SEI) layer formed on the cathode surface remains lower in NMFCO, when compared with NMFO, all the way to 200 charge cycles. Finally, we discuss possible factors affecting the changes in morphology of the cathode at the electrode-electrolyte interface, and their possible role in the improved cyclability observed with NMFCO cathodes.

NMFCO is related to P2-type $Na_{0.67}Mn_{0.65}Fe_{0.35}O_2$ (NMFO), which undergoes a structural transition to an OP4/Z phase when charged to 4.3 V, and to an orthorhombic *Cmcm* phase at 1.5 V discharge [14], [15], [16]. Modern uses of rechargeable batteries include applications such as in personal electronics, electric vehicles, and solar energy storage, where batteries are charged and discharged on a routine basis. In such situations, repeated crystal structure changes during charge cycling can severely shorten the life expectancy of a cell due to stresses on the integrity of the battery cathode.

Layered cathode materials, designed with transition-metal based crystal structures of the type described here, are known to suffer from crystal structure distortions such as the Jahn-Teller distortion [17], [18]. Often, the state of degeneracy of an electronic energy state in a given molecular system can lead to a spontaneous structural distortion to a lower energy state, typically with lower symmetry, and leading to changes in bond length [19]. Such lowering of symmetry can result in crystal structure changes during intercalation and de-intercalation of the Li or Na ions. Cobalt is known to possess a low-spin cobalt (II) oxidation state that is strongly Jahn-Teller active [20]. However, the 3+ or 4+ oxidation states of Co, generally found at higher voltage, are not Jahn-Teller active [21]. The NMFCO structure we report in this article is based on the idea that Co could help minimize structural distortions in NMFO [22].

## 2. Methods

Stoichiometric amounts of $Na_2CO_3$ (98% Alfa Aesar), $Fe_2O_3$ (99.99% Alfa Aesar), $MnO_2$ (99.9% Alfa Aesar), and CoO were reacted using a solid-state route and subsequently pelletized. Pellets were then stored inside a glovebox with an ultra-high purity Argon atmosphere. X-ray Diffraction (XRD) measurements were taken using the coupled θ-2θ mode on a Bruker D8 Diffractometer using Cu Kα radiation (λ = 1.54 Å) and a high-sensitivity Lynxeye detector. The annealed pellets were crushed and used to prepare slurries, with an 8:1:1 weight ratio of active material (NMFO or NMFCO), carbon black (C65), and polyvinylidene fluoride (PVDF) in N-methyl-pyrrolidone (NMP, 99.5% Sigma Aldrich). Coating these slurries on aluminum foil using a doctor blade produced cathodes with a loading mass of 4-5 mgcm$^{-2}$; these were used as current collectors in the coin cells. Care was taken to ensure that sheets of cathode material are of near-uniform thickness and are consistent from cell to cell. The cathodes were dried in a vacuum oven at 60°C for one day before being pressed into cells. The electrolyte used was a 1 M solution of sodium perchlorate ($NaClO_4$, 99.8% Sigma Aldrich) dissolved in a 1:1 by volume mixture of ethylene carbonate (EC) and diethylcarbonate (DEC).

Electrochemical performance of NMFCO cathodes was analyzed using assembled CR2032-type coin cells with Na as the negative electrode. The synthesis procedure followed to synthesize active material of NMFCO and coin cells is the same as the procedure followed for the synthesis of NMFO and NMFNO (the subject of a concurrent study). Coins cells were assembled and pressed in an ultra-high purity (UHP) argon-filled glovebox with water and oxygen content kept below 1 ppm. Cathodes were cycled in the voltage ranges 1.5-4.3 V, 1.5-4.0 V, and 2.0-4.0 V. We performed XRD on selected cathodes before cycling.

To study structural transitions during cycling, we performed XRD after 10 cycles on cathodes stopped at different voltages during the charge-discharge cycle. We also performed XRD on the cathode following 200 full charge-discharge cycles. For XRD, the cathodes were removed from their coin cells in a UHP Argon-filled glovebox, washed with Dimethyl Carbonate (DMC) solution to remove the electrolyte from the surface, and soaked in DMC solution for 24 hours. They were then dried at 50°C in a vacuum chamber and subsequently sealed in Argon within an airtight specimen holder specially designed, in-house, for XRD measurements. The entire process was followed without ever taking the samples out of the inert glovebox atmosphere. Our process completely avoided any exposure to air during the transfer or during the collection of XRD data.

We performed Scanning Electron Microscopy (SEM) on a JEOL JSM-6460 LV scanning electron microscope with Energy Dispersive Spectroscopy (EDS), in order to study the morphology of uncycled and cycled cathodes. Electrochemical Impedance Spectroscopy (EIS) was performed using an AutoLab PGSTAT30 potentiostat from Metrohm equipped with Nova 1.7 software. EIS measurements reported here were performed using an AC excitation signal of 10 mV and measured in the 1000 kHz to 0.1 Hz frequency range, with logarithmic frequency steps. Equivalent circuit modeling was performed using a Python program developed in-house, as described further in the discussion section and in Supplementary Materials.

## 3. Results and Discussion:

### 3.1. Rietveld analysis of X-ray diffraction (XRD) data

Our XRD studies confirm an undistorted P2 crystal structure (space group $P6_3/mmc$) for our starting cathode, $Na_{0.67}Mn_{0.625}Fe_{0.25}Co_{0.125}O_2$, with no impurity peaks or secondary phases detected in the diffraction patterns. The P2 structure is described as a stack of edge-sharing $MO_6$ layers (where M is a transition metal atom) accommodating two different prismatic sodium sites $Na_e$ and $Na_f$ which share edges and faces, respectively, with the $MO_6$ octahedra [23]. This is shown in Figure 1.

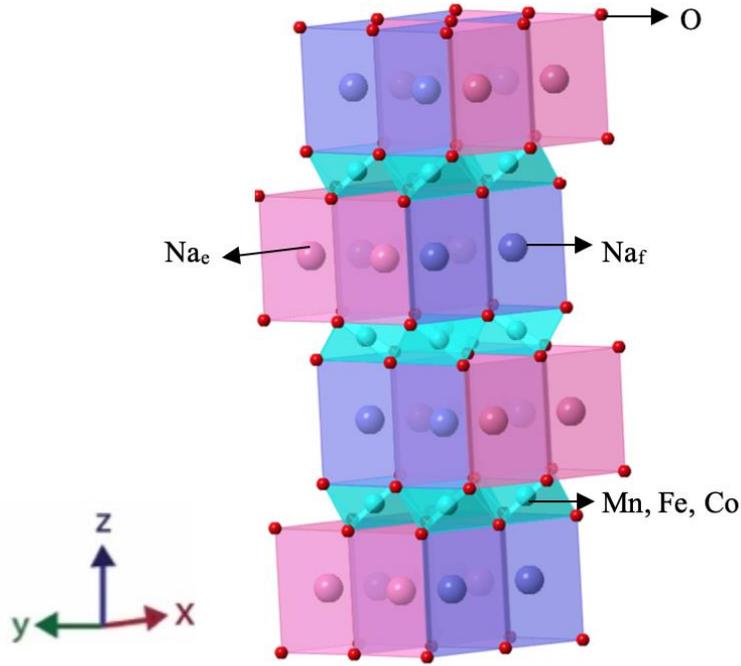

**Figure 1: Crystal structure of pristine uncycled NMFCO.** The subscripts on $Na_e$ and $Na_f$ refer to Na sites which are edge-sharing and face-sharing with the neighboring octahedra.

Figure 2a shows the XRD of uncycled and cycled NMFCO cathode material taken with the cycling stopped at different voltages in the charge-discharge cycle (also see Table 1). During the first 10 cycles, the cells were charged to 4.3 V and discharged to 1.5 V. On the 11$^{th}$ cycle, their cycling was stopped at various charge and discharge voltages. In NMFO, we observe a structural transition to a Z (also called OP4) phase at 4.3 V, and to a *Cmcm* phase at 1.5 V, as previously reported in these structures [17], [24]. These transitions are notably absent in our Co-substituted

NMFCO. We further confirm, from Rietveld refinement, that NMFCO maintains its pristine $P6_3/mmc$ crystal phase when charged to 4.3 V. New peaks are observed near the (102), (103), and (104) peaks in NMFCO at 1.5 V discharge. We interpret this as the formation of a mixture of two phases of the same crystal structure and symmetry, but with different crystalline axes, a = b in both cases. Figure 2b shows our Rietveld refinement of this two-phase mixture. The additional peaks, which appear as shoulders, are fitted with another $P6_3/mmc$ phase, with a' = b' = 3.166 Å. We observe a reversible impact on crystalline quality at high (4.3 V) and low (1.5 V) voltage in NMFCO as the XRD peaks broaden even though the crystal structure transitions are suppressed. As discussed in subsequent sections, the suppression of such crystal structure transitions contributes significantly to the state of health (SoH) and cyclability of the cells cycled up to 200 cycles.

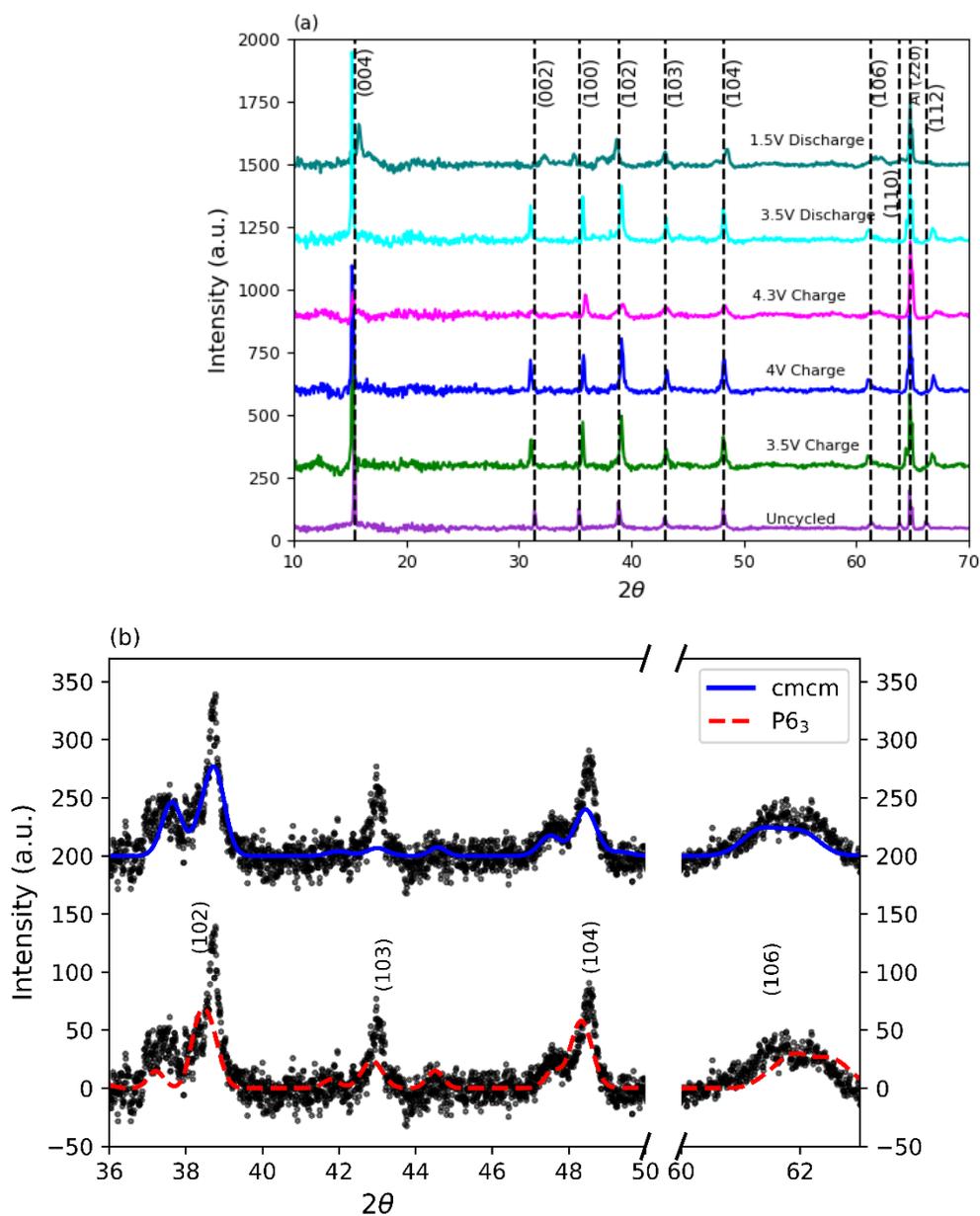

**Figure 2: X-ray Diffractograms (XRD) of NMFCO cathode at different charge or discharge voltages.** (a) XRD of NMFCO cathode material charged and discharged at different voltages (also see Table 1) during cycling in the voltage range 1.5–4.3 V (b) XRD of NMFCO at 1.5 V showing fits to *Cmcm* and *P6$_3$/mmc* phases.

| Crystal lattice parameters from Rietveld refinement of $Na_{0.67}Mn_{0.625}Fe_{0.25}Co_{0.125}O_2$ during cycling in the 1.5-4.3 V range | | | |
|---|---|---|---|
| **Voltage (V)** | **a = b (Å)** | **c (Å)** | **Phase** |
| Uncycled | 2.900 | 11.329 | $P6_3/mmc$ |
| 3 V Charge | 2.886 | 11.349 | $P6_3/mmc$ |
| 4 V Charge | 2.875 | 11.402 | $P6_3/mmc$ |
| 4.3 V Charge | 2.856 | 11.399 | $P6_3/mmc$ |
| 3.5 V Discharge | 2.877 | 11.392 | $P6_3/mmc$ |
| 1.5 V Discharge | 2.969 | 11.058 | $P6_3/mmc$ |
| | 3.166 | 10.869 | |

**Table 1.** Evolution of lattice parameters of P2-type NMFCO at different voltages during cycling in the 1.5–4.3 V range. Shown here are results from Rietveld refinement of the data displayed in Figure 1. As indicated in the text, the 1.5 V discharge data was also fitted to a *Cmcm* phase for comparison [see Supplementary Materials].

Table 1 shows results from Rietveld analysis of the XRD diffractograms shown in Figure 2a. Note changes in lattice parameters of NMFCO at different voltages during cycling in the 1.5-4.3V range. These changes show up as plateaus in the galvanostatic cycling studies described in the following section [also see Figure 5a].

With increasing charging voltage, the c-axis increases due to de-intercalation of Na ions from prismatic sites. Intuitively, one might expect the c-axis to decrease with Na de-intercalation – however, it is now understood that de-intercalation of Na ions results in a decrease of the *screening effect* provided by Na ions [24], [25]. This increases the repulsive interaction between oxygen layers, thus leading to an expansion of c-axis. The in-plane a and b axis decrease at higher voltages due to the oxidation of transition elements sitting inside the $MO_6$ octahedron [24], [26]. The appearance of a sloping potential profile with a plateau at high voltages is consistent with a low degree of c-axis distortion with the insertion of ions into the lattice [25]. We have fitted the XRD data at 1.5 V discharge using two $P6_3/mmc$ phases with different a- and b-axis lattice parameters. This gives a strong fit, but we recognize that previous researchers studying similar cathodes in the NMFO family have fitted the low-voltage data with a *Cmcm* phase [28]. We have attempted this fit for comparison, and have found that it gives us similar fit parameters to the fit with two $P6_3/mmc$ phases. Therefore, we present both fits in Figure 2b.

| Crystal lattice parameters from Rietveld refinement of $Na_{0.67}Mn_{0.625}Fe_{0.25}Co_{0.125}O_2$ after 200 cycles in 1.5-4.0 V and 2.0-4.0 V range | | | |
|---|---|---|---|
| **Voltage Range (V)** | **a = b (Å)** | **c (Å)** | **Phase** |
| Uncycled | 2.900 | 11.329 | $P6_3/mmc$ |
| 1.5-4.0 | 2.914 | 11.293 | $P6_3/mmc$ |
| 2.0-4.0 | 2.929 | 11.101 | $P6_3/mmc$ |

**Table 2.** Crystal lattice parameters of uncycled P2-type $Na_{0.67}Mn_{0.625}Fe_{0.25}Co_{0.125}O_2$ after 200 cycles in the 1.5–4.0 V and 2.0–4.0 V ranges. These parameters correspond to Rietveld analysis of XRD data shown in Figure 5.

Table 2 shows the crystal lattice parameters of uncycled P2-type NMFCO, together with results after 200 cycles, in the 1.5–4.0 V and 2.0–4.0 V ranges. This uses data from the diffractograms in Figure 3, where the cycled cathodes are scanned at the open-circuit voltage. It is important to note that although we find a suppression of mid-cycle crystal structure changes in NMFCO when compared with NMFO, both cathode types revert to their pristine P2-type crystal structure in the open-circuit state following 200 cycles.

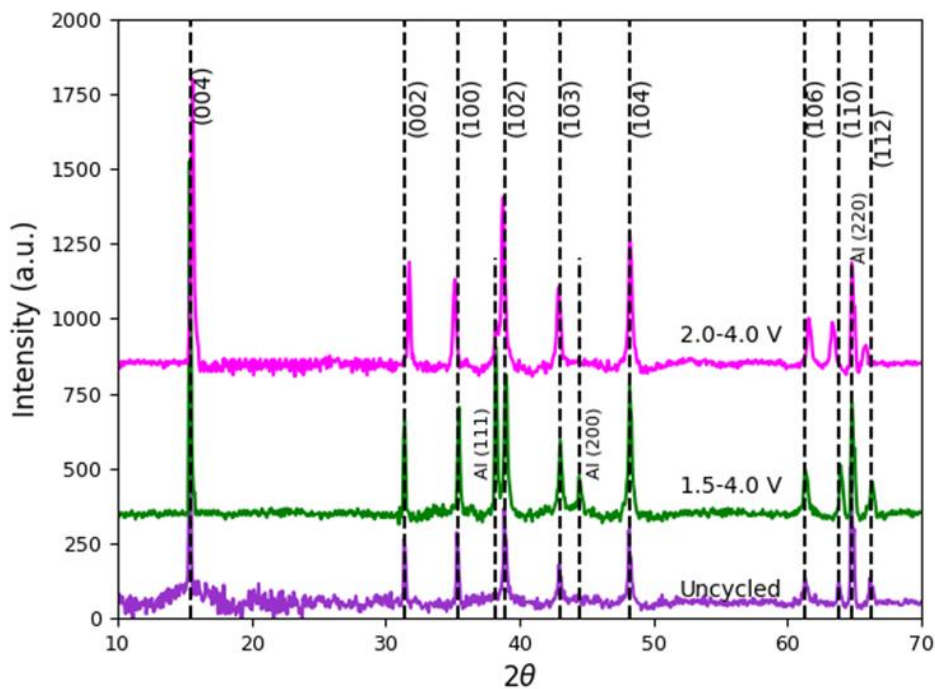

**Figure 3. XRD diffractogram of uncycled and cycled NMFCO.** XRD of $Na_{0.67}Mn_{0.625}Fe_{0.25}Co_{0.125}O_2$ cathode surface after 200 cycles in the 1.5–4.0 V and 2.0–4.0 V ranges. Diffraction peaks for Aluminum 111 and 200 are showing in the 1.5-4.0V range, probably because a portion of the Aluminum current collector was (unintentionally) exposed during that XRD measurement.

### 3.2. Galvanostatic Cycling & Battery Performance

Next, we examine the impact of Co substitution on cycling performance, discharge capacity and specific energy of cathodes. Figure 4 shows the discharge capacity and specific energy of NMFCO and NMFO cathodes during constant current charging and discharging. The uniform size and thickness of the cathodes ensures that they have the same form factor and can be compared using these measurements. We cycled NMFCO and NMFO up to 100-200 times in the 1.5–4.3 V, 1.5–4.0 V, and 2.0–4.0 V voltage ranges. In each voltage range, the first two formation cycles are at C/20 rate (0.047 mA) and the remaining cycles are at C/10 rate (0.0945 mA). These voltage ranges were chosen based on voltages at which structural changes were observed in NMFO during charge-discharge cycling, with an intent to examine how the suppression of crystal structural transitions in NMFCO contributes to electrochemical properties during cycling [24]. Indeed, we observe a significant advantage in NMFCO over NMFO.

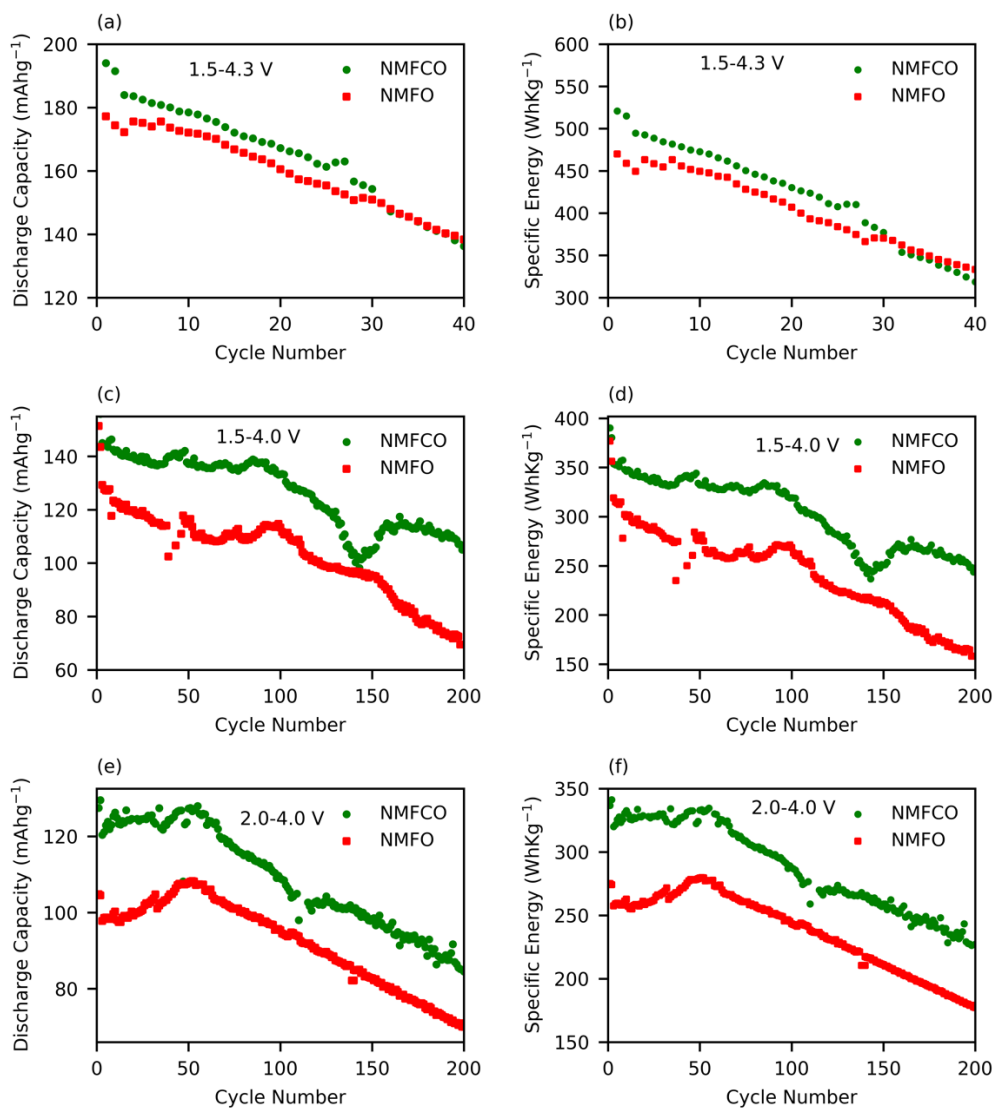

**Figure 4. Discharge Capacity and Specific Energy of NMFO and NMFCO.** Shown here are measurements in voltage ranges (a, b) 1.5–4.3 V; (c, d) 1.5–4.0 V; and (e, f) 2.0–4.0 V.

In the 1.5-4.3 V range, NMFCO displays a first-cycle specific energy of 520 Whg$^{-1}$ and a capacity of 195 mAhg$^{-1}$, when compared with 470 Whg$^{-1}$ and 177 mAhg$^{-1}$ for NMFO. In this range, we display data only up to 40 cycles in Figure 4. This is because our values for discharge capacity showed unusually large fluctuations, possibly due to electrolyte degradation in these cells, and potentially indicating the need for further work in electrolyte chemistry for extended cycling in this range. In the 1.5-4.0 V range, a higher discharge capacity and specific energy are retained by NMFCO up to 200 cycles. In particular, NMFCO retains 71% capacity after 200 cycles as compared to 55% capacity in NMFO. In the 2.0-4.0 V range, NMFCO retains 66% capacity after 200 cycles while NMFO retains 65% capacity. Thus, we conclude that there is an improvement in discharge capacity and voltage cycling with NMFCO when compared with NMFO in all three voltage cycling ranges (1.5–4.3 V, 1.5–4.0 V, and 2.0–4.0 V). Liu et al assert that $Na_{2/3}Mn_{1/2}Fe_{1/4}Co_{1/4}O_2$ displays higher energy capacity than other layered cathodes in the P2-type family of cathodes known at the time. While our results for NMFCO display similarly high energy capacity, our extended cycling results show that the discharge capacity of $Na_{0.67}Mn_{0.625}Fe_{0.25}Co_{0.125}O_2$ is nearly equal to P2-$Na_{0.67}Mn_{0.625}Fe_{0.25}Ni_{0.125}O_2$. From these results, it is clear that our low-cobalt composition shows promise for high capacity and relatively good cyclability. At this early stage in the study of sodium-ion batteries, it is important to explore different stoichiometries with different active materials, and our work extends previously fruitful study into cobalt-containing NMFO-group cathodes.

Figure 5 shows galvanostatic charge cycling curves with NMFCO as cathode, cycled in the same voltage ranges (1.5–4.3 V, 1.5–4.0 V, and 2.0–4.0 V), as described in Figure 4. Note that the plateaus observed near 3.75 V in the 1.5-4.3 V cycled voltage range in Figure 5a are absent in the 1.5–4.0 V (Figure 5b) and 2.0–4.0 V (Figure 5c) cycled voltage ranges. As stated in the previous section, the plateaus observed in the 1.5-4.3 V range are consistent with the insertion of ions into the lattice [25]. It should be noted that plateaus at high voltage in other materials have also been found to be attributable to the filling of oxygen and transition metal vacancies in the bulk by ions migrating from the surface, reactions stemming from degradation of the electrolyte, and the nature of the redox reaction when cycling at higher voltage [29], [30], [31], [32].

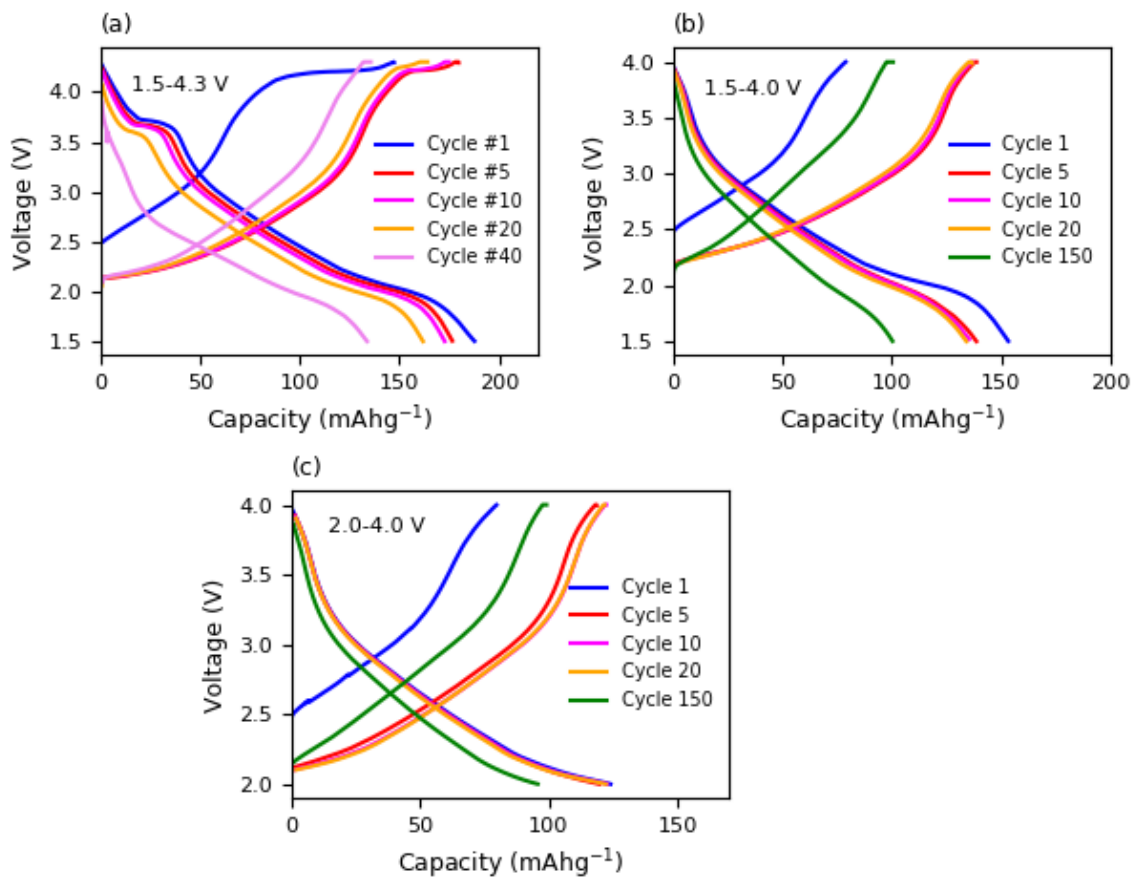

**Figure 5. Galvanostatic cycling of NMFCO cathodes.** Cycling curves of NMFCO cathodes cycled in the voltage ranges 1.5-4.3 V, 1.5-4.0 V and 2.0-4.0 V.

### 3.3. Electrode-electrolyte interface and internal resistance

To obtain more information about the internal resistance of the cell, we looked at the initial voltage drop in the galvanostatic charge-discharge cycling plots. An equivalent series resistance can be calculated from the voltage drop during the reversal of polarity [33]. This can be characterized as the internal resistance of the cell [34]. This is an important quantity to isolate, because it can be related to the degradation of the electrolyte, which causes more rapid capacity fading and thus poor cyclability. In the galvanostatic plots, the initial voltage drop is obtained by finding the initial and final voltages in the region where the drop is linear. This is divided by the constant applied current to obtain the equivalent resistance [33].

We now use the initial slope from galvanostatic results shown in Figure 5 to compare the internal resistance of the cells during charge-discharge cycling [35]. This initial slope characterizes active polarization at the cathode surface, which is in turn driven by Charge Transfer processes at the electrode-electrolyte interface [35], [36]. Figure 6 shows the initial slope of galvanostatic discharge curves plotted versus cycle number for NMFO and NMFCO cathodes cycled under the same conditions. We find that general trends between NMFO and NMFCO do not significantly change for other reasonable choices of energy capacity range. The linear region used to find the initial slope was chosen to be in the 0 - 3.6mAhg$^{-1}$ energy capacity range. Extreme outliers, attributed to voltage fluctuations, are not shown in Figure 6.

An inspection of the initial slope of NMFCO shown in Figure 6 immediately reveals that the internal resistance of NMFCO is less as compared with NMFO (also see our previous data [14].) We attribute this to an improvement in kinetic hindrance to the charge transfer process at the cathode-electrolyte interface [35]. Such kinetic hindrance to charge transfer typically presents itself as an increase of the total internal resistance of the cell, as also demonstrated by our analysis and modeling of the Electrochemical Impedance Spectroscopy discussed in later sections (see, for example, Table 3). Figure 6 thus demonstrates that NMFCO cells start off with relatively low internal resistance when compared with NMFO and continue to have lower internal resistance with cycling.

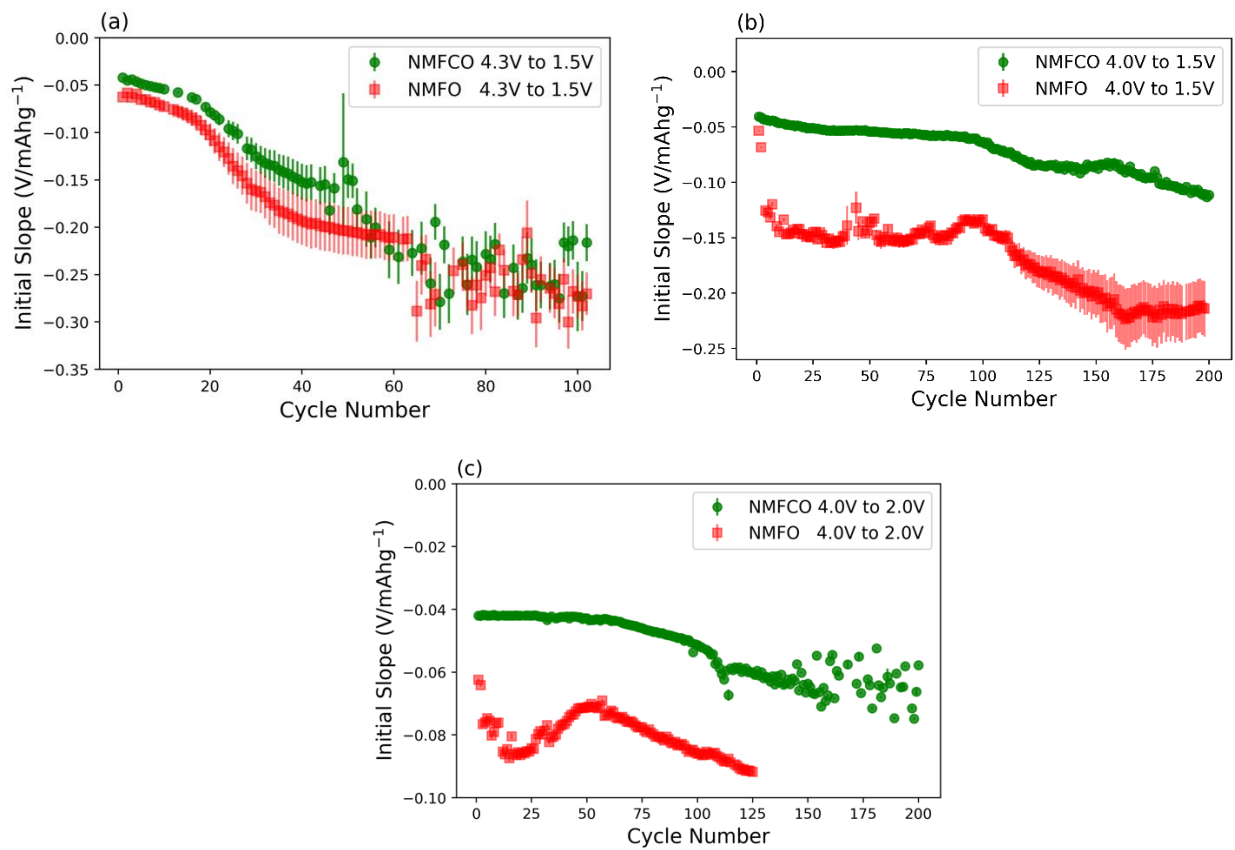

**Figure 6. Initial slope of galvanostatic discharge curves as a function of cycle number.** The data was obtained from our galvanostatic charge-discharge measurements, including data shown in Figure 3 for different cycle numbers, for NMFO and NMFCO cathodes in the (a) 1.5-4.3 V, (b) 1.5-4.0 V and (c) 2.0-4.0 V ranges. Points depict best fit slopes in the 0-3.6 mAhg$^{-1}$ energy capacity range.

Note that the size of the error bars in Figure 6 increases beyond a certain number of cycles (beyond 20 cycles in Figure 6a, and beyond 100 cycles in Figure 6b). This is due to a deviation from linearity of the initial slope of the galvanostatic curve, indicating a change in nature of the resistive element of the equivalent circuit shown in Figure 8. When compared with NMFCO cells, the error bars in NMFO increase early in their respective cycling histories. This implies that, with repeated cycling, the interface at the NMFO cathode degrades earlier than NMFCO. The effect of this on the equivalent circuit is described in a future publication [37]. We conclude, based on Figure 6, that the active polarization and equivalent internal resistance of NMFCO is less than that of NMFO. This implies a superior state of health of our NMFCO-based cells when compared with our NMFO cells.

### 3.4. Scanning Electron Microscopy

To further investigate the cause of capacity degradation in NMFCO cathodes, we performed SEM on NMFCO cathodes before and after cycling in the voltage range 1.5-4.0 V. The electron micrographs are shown in Figure 7. We observe cracks on cathode surfaces cycled in the 1.5-4.0 V and 2.0-4.0 V range. This is an aspect of battery electrode health that has not been extensively studied in Na-based cathodes. We argue intuitively that cracks on the cathode surface could arise due to the high internal strain developed with repeated cycling during sodium intercalation and de-intercalation into the crystal lattice. The volume of unit cell decreases from 95.28 $\text{Å}^3$ for uncycled cells to 93.91 $\text{Å}^3$ for cells charged to 4 V. Volume change in the unit cell due to intercalation and deintercalation of Na ions could lead to crack formation even in the absence of structural changes. Additionally, a mismatch in the shared surface area between the SEI and the cathode due to the expansion of the cathode could promote cracking in the SEI layer.

In order to quantify the morphology of cracks observed after cycling in our cathode surfaces, we developed a program that calculates the percentage of a sample surface occupied by cracks as seen in an SEM image. Averaging over 16 images of small (800 micron by 800 micron) areas of the cathode surface on samples of NMFO and NMFCO, we find that the percentage of the surface showing cracks in NMFO is approximately 5.5%, which is 40% higher than the 3.9% found for NMFCO. A lower percentage of cracks would also lead to a decrease in interfacial resistance and, consequently, an improvement in discharge capacity [38], [39]. The smaller percentage of cracks observed in NMFCO agrees with the conclusions from Figure 6 (section 3.3) and Table 3 (section 3.6), which imply that interfacial resistance is lower in NMFCO.

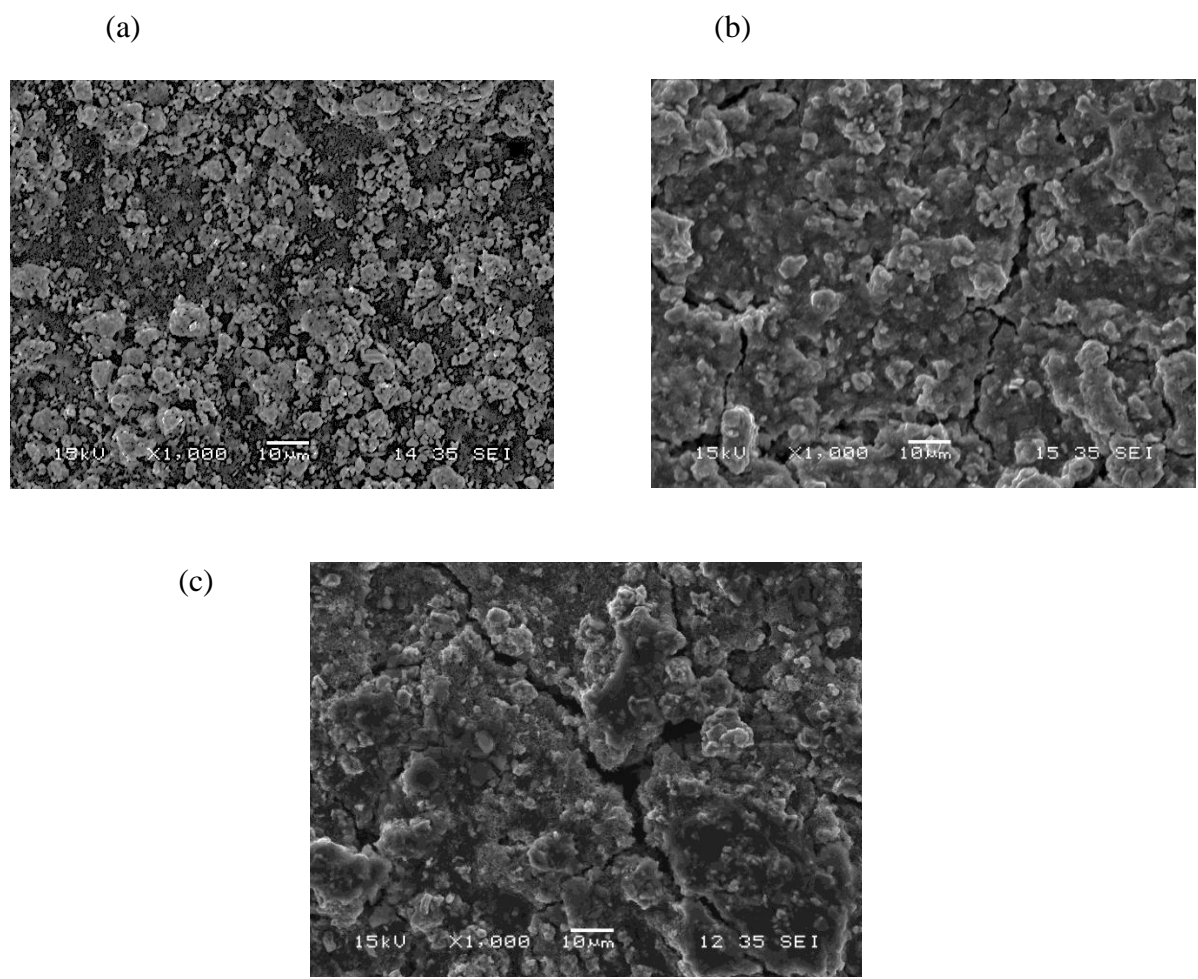

**Figure 7. Scanning Electron Micrography (SEM).** SEM images of $Na_{0.67}Mn_{0.625}Fe_{0.25}Co_{0.125}O_2$ (NMFCO) cathode surfaces as prepared, and after cycling. (a) Uncycled cathode; (b) after 200 cycles in the 1.5–4.0 V range; (c) after 200 cycles in the 1.5–4.0 V range. We conclude that formation of a Solid Electrolyte Interface (SEI) is observed in 'b' and 'c'.

The formation of surface cracks can also be affected by other materials parameters and morphology such as the type of binder, grain size, the type of electrolyte, and the rate of formation of the electrolyte interface. Electrolyte consumption increases with increasing crack percentage, implying that higher crack percentages lead to higher consumption of electrolyte, shorter cycle life, and degraded capacity [39], [40]. Note that the texture of the uncycled NMFCO cathode is grainy, as seen in Figure 7a. In contrast, the post-cycled surface texture is not as grainy (Figures 7b and 7c). This matte-like deposition on the surface indicates the formation of a Solid Electrolyte Interface (SEI) layer, formed due to chemical/electrochemical reactions occurring at the electrolyte-electrode interface. Electrolyte depletion occurs as the result of electrolyte penetration through cracks, leading to further reactions and growth of the SEI layer. The growth of an SEI layer within cracked surfaces helps improve contact with the electrolyte while depleting the electrolyte at the same time. As described below, the morphology of the SEI layer is intimately connected with both internal resistance and the high-frequency electrochemical impedance response. We present here Electrochemical Impedance Spectroscopy (EIS) to gain further insight into the nature of this interface.

### 3.5. Electrochemical Impedance Spectroscopy

EIS was performed on cells of NMFO and NMFCO without cycling and after 10 and 200 cycles. Our results and analysis are shown in Figures 8 (a)-(c). For both NMFO and NMFCO, the same cells were used for the uncycled and 10 cycle tests; a different cell was used for the 200 cycle test. Figure 8a shows Nyquist plots, displaying the out-of-phase component of the complex, frequency-dependent impedance of each cell as a function of its in-phase component. Note that higher frequencies appear near the origin of such a Nyquist plot. The highest frequency semicircle in a Nyquist plot is typically associated with the formation of an SEI layer. In our results, we note a depressed semicircle near the origin of our Nyquist plots. This becomes more prominent for both types of cells with increased cycling [41], [42]. Semicircles roughly in the 1 kHz – 10 mHz frequency range in similar cathode materials are thought to correspond to charge-transfer processes from the electrolyte to the SEI and from the SEI to the cathode. Low frequency is typically characterized by diffusion processes in the electrodes [43]. Figure 8b is illustrative of the difference in behavior between NMFO and NMFCO at low frequencies as cycling is increased. Figure 8b shows the total impedance plotted against the logarithm of frequency (decreasing to the right, to correspond to the Nyquist plot). We see that the total impedance of NMFO and NMFCO is similar at higher frequencies. However, it becomes dissimilar at higher cycle numbers and at lower frequencies. This is associated primarily with charge-transfer processes as discussed further in the next section.

(a)

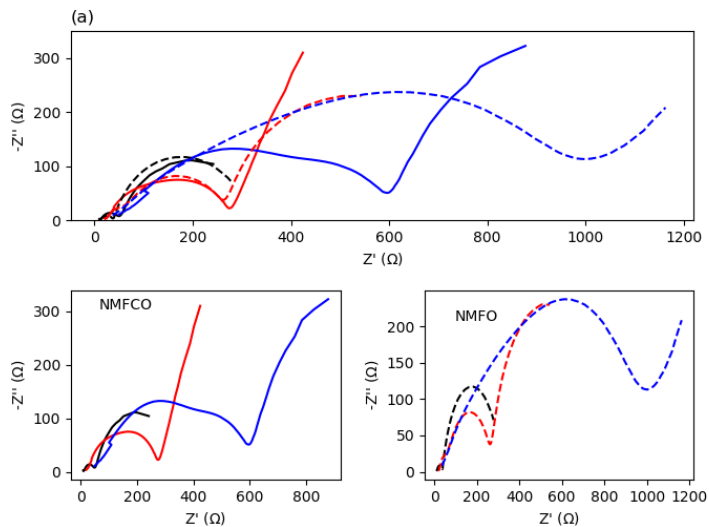

(b)

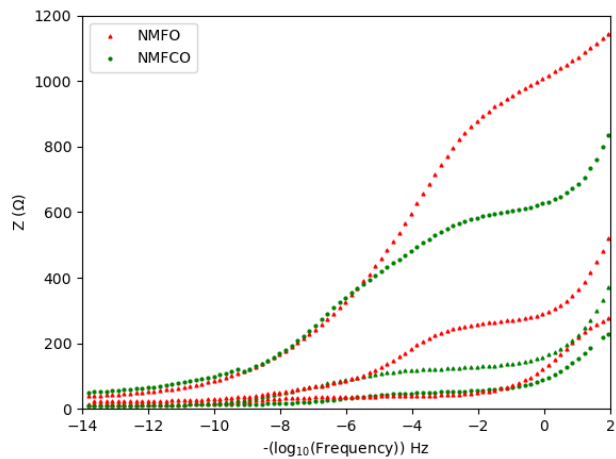

**Figure 8. Electrochemical Impedance Spectroscopy results.** (a) Nyquist plots of NMFO (dashed line) and NMFCO (solid line). Data is shown for uncycled cells (black), and for cells after 10 cycles (red) and 200 cycles (blue) cycled in the 2–4 V range. All the spectra are obtained by varying frequency from 1000 kHz to 0.1 Hz. (b) Total impedance of uncycled and cycled (10 cycles, 200 cycles) cells of NMFO (red triangle) and NMFCO (green circle) as a function of negative log of frequency.

### 3.6. Modeling of equivalent circuits from Electrochemical Impedance Spectroscopy

To further understand the electrochemistry and the State of Health (SoH) of our cells, we have modeled equivalent circuits for the cells studied by EIS. For this work, our group developed an in-house equivalent circuit modeling program in Python, described in the Supplementary Materials. We consider and compare two equivalent circuit models: Model A shown in Figure 9a and Model B shown in Figure 9b. Model A is informed by the shape of the measured EIS spectra [44], [45], whereas model B is informed by the dominant physical processes expected to occur in a half cell [46], [47].

(a)

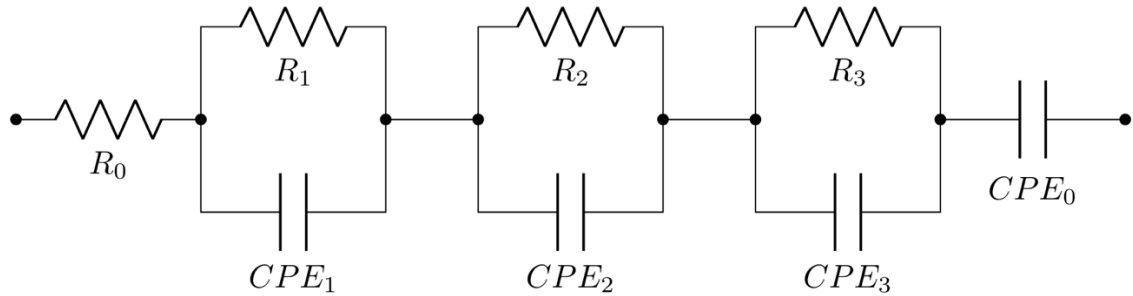

(b)

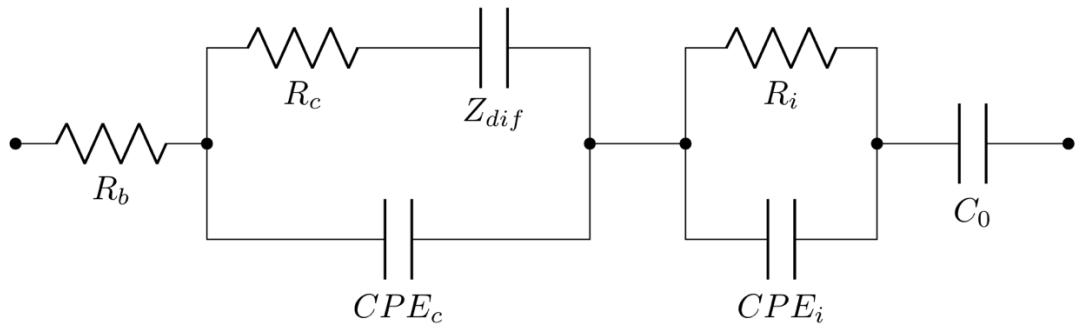

**Figure 9. Equivalent Circuit Fitting from Electrochemical Impedance Spectroscopy.** (a) Model A: Equivalent circuit diagram obtained after analysis of impedance spectra of cycled and uncycled NMFCO and NMFO half cells. (b) Model B: Equivalent circuit diagram describing a half cell with series bulk resistance and capacitance, diffusion of sodium ions inside the cathode, and charge transfer and double layer contributions from the cathode and interphase layer.

For model A we begin with an examination of the impedance spectrum to determine the number and type of the circuit elements that are needed to describe the measured EIS spectra. Starting with an initial estimate of the values for each circuit element based on the size and shape of the features in the Nyquist plot, the program generates a best fit for the circuit elements in the model. A resistor, three pairs of a resistor in parallel with a constant-phase element (CPE), and a separate CPE effectively modeled our cells. The simulated dataset agreed well with the measured real and imaginary components of impedance versus frequency data for each cell.

A CPE is used in equivalent circuits to model the behavior of a double layer. It describes an "imperfect" capacitor that does not have a fixed boundary due to the porosity of the material. The impedance of a CPE, Z, can be described by $1/Z = Q(\omega i)^n$. Here, Q is the analog of capacitance for the CPE, with the important difference that its units of Q are not Farad (F = $s\Omega^{-1}$), but $s^n\Omega^{-1}$, and n is the phase. In Tables 3 and 4, resistances are given in $\Omega$, Q in $s^n\Omega^{-1}$, and n is dimensionless. The phase of the impedance of a CPE is given by $-(90*n)°$. The bounds of $n$ are $0 \leq n \leq 1$, where $n = 0$ describes purely resistive behavior and $n = 1$ represents purely capacitive behavior. A CPE and a resistor in parallel yields a semicircular feature in the Nyquist plot, while a CPE by itself gives a straight line with a slope characterized by its phase. In some cases, a semicircle, or the straight-line "tail" of the Nyquist plot at low frequencies, could not both be fitted reliably. In such cases the resistor-CPE loop and/or the separate CPE are excluded from the fit, and their parameters are left blank in Table 3. Due to this difference in equivalent circuit elements used to describe the low frequency region of the Nyquist plot, it is difficult to interpret the low frequency best fit values across different batteries and different cycles.

Model B, described in Figure 9b, describes the dominant electrochemical processes expected to occur in a half cell. The resistor connected in series models bulk resistance in the cell. A resistor and CPE connected in parallel describes an imperfect leaked capacitor which can model the electrolyte interphase. A Randles-like circuit with a resistor and diffusion impedance connected in parallel with a CPE which models charge transfer, diffusion, and an electric double layer in the cathode. Finally, a capacitor is added in series to account for any capacitance which may show up in the measurement process. From Fick's second law, diffusion impedance through an electrode bounded semi-infinite block can be written as $Z_{dif} = A(i\omega)^{-n}\tanh(B(i\omega)^n)$ where n=0.5. In the high frequency limit, this expression reduces to the classic Warburg element. However, to allow for non-ideal diffusion behavior, we let n be a parameter in our fit and use the full expression for diffusion.

We have examined the variation of the best fit equivalent circuit parameters with cycling to gain insight into the corresponding physical processes of the cell. We perform equivalent circuit modeling with model A and B for NMFCO and NMFO uncycled, 10-cycled, and 200-cycled cells. The data for 200 cycles is from a separate set of cells which were synthesized, assembled, and cycled identically to the other cells shown here. All fitting was done in the 26.7 kHz to 100 mHz frequency range. Average residuals (Avg Resid) is calculated by

$\frac{1}{N}\Sigma \sqrt{\left(Z_{Im,fit} - Z_{Im,measured}\right)^2 + \left(Z_{Re,fit} - Z_{Re,measured}\right)^2}$, where N is the number of frequencies where measurements were taken, and Re and Im denote the real and imaginary component of impedance respectively. Table 3 shows the best fit resistor and CPE values for model A (Figure 9a). Table 4 shows the best fit parameters for model B (Figure 9b).

| Cathode | Cycle | $R_0$ | $Q_0$ $*10^{-3}$ | $n_0$ | $R_1$ | $Q_1$ $*10^{-6}$ | $n_1$ | $R_2$ | $Q_2$ $*10^{-6}$ | $n_2$ | $R_3$ | $Q_3$ $*10^{-3}$ | $n_3$ | Avg Resid |
|---|---|---|---|---|---|---|---|---|---|---|---|---|---|---|
| NMFCO | 0 | 13 | -- | -- | -- | -- | -- | 39 | 86 | 0.75 | 274 | 3.8 | 0.86 | 2.6 |
| | 10 | 23 | 3.8 | 0.77 | 151 | 13 | 0.72 | 106 | 21 | 0.89 | -- | -- | -- | 3.1 |
| | 200 | 58 | 3.4 | 0.65 | 430 | 9.6 | 0.66 | 93 | 45 | 1.0 | -- | -- | -- | 7.5 |
| NMFO | 0 | 10 | -- | -- | -- | -- | -- | 30 | 43 | 0.63 | 275 | 1.7 | 0.89 | 1.1 |
| | 10 | 24 | -- | -- | 46 | 14 | 0.76 | 197 | 35 | 0.85 | 516 | 2.8 | 0.93 | 1.6 |
| | 200 | 32 | 7.4 | 0.70 | 658 | 56 | 0.48 | 319 | 45 | 0.82 | 111 | 6.1 | 0.69 | 1.1 |

**Table 3: Model A** best fit values for R, Q, and n. Resistances are given in $\Omega$, $Q_0$ in $s^n\Omega^{-1}$, and n is dimensionless. The R values reported for Resistor-CPE loops are given by the fit, although the semicircle in the Nyquist plot is not completed in the measured frequency range in some cases. Subscripts 1, 2 and 3 in the top row refer to highest (1), mid (2), and lowest (3) characteristic frequencies.

| Cathode | Cycle | $R_b$ | $R_i$ | $Q_i$ *$10^{-6}$ | $n_i$ | $R_c$ | $Q_c$ *$10^{-6}$ | $n_c$ | $A_d$ | $B_d$ | $n_d$ | $C_0$ *$10^{-3}$ | Avg Resid |
|---|---|---|---|---|---|---|---|---|---|---|---|---|---|
| NMFCO | 0 | 11 | 8 | 23 | 0.76 | 25 | 16 | 1.0 | 137 | 1.2 | 0.55 | 27 | 3.7 |
| | 10 | 21 | 127 | 8.5 | 0.76 | 123 | 14 | 0.92 | 144 | 1.9 | 0.54 | 7 | 2.0 |
| | 200 | 62 | 400 | 9.1 | 0.67 | 84 | 68 | 0.82 | 199 | 1.6 | 0.55 | 9 | 6.9 |
| NMFO | 0 | 6 | -- | -- | -- | 26 | 5.7 | 0.79 | 250 | 0.94 | 0.55 | 49 | 4.8 |
| | 10 | 22 | 41 | 1.5 | 0.97 | 187 | 24 | 0.89 | 220 | 1.5 | 0.57 | 21 | 5.1 |
| | 200 | 26 | 680 | 53 | 0.48 | 275 | 25 | 0.93 | 178 | 1.14 | 0.41 | 11 | 4.6 |

**Table 4: Model B** best fit values for R, Q, A, B, and n. Resistances are given in $\Omega$, Q and 1/A in $s^n\Omega^{-1}$, B in $s^n$, and n is dimensionless. Subscript b corresponds to bulk processes, subscript i corresponds to interphase processes (identified by a high-characteristic frequency), subscript c corresponds to charge transfer and double layer in the cathode (identified by a mid-characteristic frequency), and subscript d corresponds to diffusion (occurring at low frequency).

In Table S1 and S2 [in Supplementary Materials], we give the effective capacitance, $C_{eff}$, and characteristic time constants $\tau = RC_{eff}$ for the resistor-CPE combinations. Our calculations are based on the equation given by Hsu and Mansfeld, and others, $Z_{(CPE)} = Q_0^{-1}(j\omega)^{-n}$ for a resistor CPE combination where $Q_0R = t^n$ and $t = RC$. The effective capacitance was calculated as $C_{eff} = Q_0^{1/n}R^{(1-n)/n}$, with R the fitted value for the resistor-CPE combination (the width of the semicircle) [48], [49], [50].

When a solid electrolyte interphase (SEI) forms at the boundary of the electrolyte with a cathode or an anode, the impedance behavior can be modeled by a resistor in parallel with a CPE [41], [42]. The resistor models the charge-transfer at the electrolyte interface, while the CPE models the capacitive behavior exhibited by the formation of an electric double layer. Intuitively, this can be thought of as a 'leaked' capacitor with current (charge transfer) leaking through a capacitor (electric double layer).

Previous research in Li-ion cathodes relates the high-frequency semicircles to interface layers between the cathode and the electrolyte, including SEI layers. Deng et al. and Nagasubramanian argue that the impedance from interface layers at the anode is negligible, and that the increase in cell impedance comes mostly from the cathode [41], [42], [51], [52], [53], [54]. Further, all our cells were measured under the same conditions and identical Na anodes, which makes us confident that our comparison of impedance measured from one cathode to another is meaningful. Our data, shown as resistor-CPE loop 1, characterizes the fit of the high frequency semicircle. As expected, before cycling and therefore the electrochemical formation of the interphase layer, both NMFO and NMFCO show no or very small best fit values for $R_1$ or $R_i$ in both Model A and B respectively. We observe an increase in $R_1$ or $R_i$ in both Model A and B with cycling. We interpret this as the result of SEI layer growth with cycling. In NMFCO, we observe a greater increase in this resistance in the first 10 cycles compared to NMFO. However, by cycle 200 $R_1$ or $R_i$ is significantly less for NMFCO compared to NMFO. These results hint at an interphase layer which potentially forms more quickly for NMFCO and remains more stable compared to NMFO. This is in good agreement with our observation of more surface cracking for NMFO than for NMFCO and with our observed suppression of the structural transition in NMFCO. In this scenario, structural transitions and surface cracking could lead to additional solid electrolyte interphase layers forming, resulting in an increased resistance.

Mid-frequency charge transfer kinetics is described by $R_2$ in model A and $R_c$, associated with cathode charge transfer kinetics, in model B. For NMFO we observe a significant increase in best fit resistance with increasing cycling. On the other hand, NMFCO shows an increase in resistance between uncycled and 10 cycles, and then a slight decrease in resistance by cycle 200, suggesting relatively stable charge transfer kinetics after the first 10 cycles. At both cycle 10 and 200, the best fit resistance is significantly larger for NMFO compared to NMFCO.

Internal resistance $Z'$ of a cell is generally dominated by CTK, which deteriorates progressively with cycling, resulting in an effective increase in internal resistance. The contribution from CTK ($Z_{CTK}$) is approximately given by $R_1 + R_2$ from Table 3 or $R_i + R_c$ from Table 4. We find a $Z_{CTK}$ more than 1.8 times larger for NMFO (Model A: 977, Model B: 955) than for NMFCO (Model A: 523, Model B: 484). Additionally, As seen from Table 3 and Fig. 8a, we find that the increase in resistance at the lowest measured frequency for NMFO cells, from uncycled to 200 cycles, is 15% larger when compared with NMFCO. This is further confirmed by our results shown in Figure 6, where we find that the magnitude of the initial slope of the galvanostatic plots, representing interfacial resistance, increases with cycling. Thus, NMFCO performs better than NMFO at maintaining a lower internal resistance $Z'$ up to 200 cycles.

The low-frequency impedance is usually associated with diffusion, which is generally fitted with a Warburg element (a CPE with a phase of 45°, or 0.5 in our units of n). However, we find that the low frequency impedance response cannot be fully described by a single Warburg element. In model A we use a combination of a CPE ($CPE_0$) and a resistor CPE loop ($R_3$, $CPE_3$) which leads to a good fit. However, some EIS spectra require none, some, or all of these elements, making it difficult to compare best fit parameters. On the other hand, model B effectively characterizes the low frequency impedance response, at the cost of a slightly worse average residuals for some spectra. $A_d$ scales the diffusion impedance, $Z_{dif}$, given above. We find that $A_d$ is comparable between NMFO and NMFCO at 200 cycles. In the ideal case where $n_d$=0.5, $B_d \propto \frac{L}{\sqrt{D}}$ where L is the particle length, and D is the diffusion constant [47]. We find a greater decrease in $B_d$ for NMFO compared to NMFCO between cycle 10 and cycle 200. This is consistent with the argument of structural transitions and surface cracking in NMFO, which may be leading to increasing smaller particle size with cycling.

Our cells show very good cycling performance up to 30 cycles, as also seen in other reports. Here, we have extended our study up to 200 cycles. At these extended cycles, we think that the internal resistance Z', dominated by CTK, progressively deteriorates with cycling as shown from the initial slope in Fig. 6 and from the data displayed in Table 3. It is not yet clear to us as to whether this is an inherent property of the cathode material or if this deterioration can be solved with further research into cell configuration. Another parameter to consider is surface cracking which leads to increased consumption of the electrolyte due to the formation of new SEI layers within the cracks. Our data at high cycle numbers up to 200 cycles indicates that cycling performance can degrade due to overconsumption of the electrolyte resulting from such cracks.

Increase in internal resistance with cycling is a common cause of battery failure. We tentatively conclude, subject to further work, that the chemistry at the cathode-electrolyte interface for NMFCO causes an SEI layer to form, but that this SEI layer does not dramatically deteriorate with increased cycling. Thus, the low-resistance SEI layer helps form a better contact with the electrolyte, effectively decreasing the internal resistance of the battery with NMFCO as cathode. The fact that NMFCO experiences a significantly smaller increase in total internal resistance, with cycling up to 200 cycles, suggests that this cathode is a good candidate for batteries with longer cycle-lifetime.

## 4. Conclusions

We present new results on NMFCO, a P2-type Na-ion cathode composition with Co concentration much lower than previously attempted. In coin cells with Na as an anode, we demonstrate the absence of charge cycling-driven structural distortions which are potentially a problem in other P2-type cathode material compositions such as $Na_{0.67}Mn_{0.65}Fe_{0.35}O_2$ (NMFO) [14], [24], [55]. We employ a combination of Rietveld analysis of X-ray diffraction with cycling in coin cells, morphological examination of the cathode surfaces, detailed modeling of Electrochemical Impedance Spectroscopy of cycled cells, and slope analysis of galvanostatic plots. We also report results up to 200 cycles, thus extending such studies beyond those reported in other Na-based cathodes. Our studies reveal that the new composition exhibits superior performance when compared with P2-type cathodes such as NMFO. In particular, we report the absence of a previously reported structural transition to the so-called "Z" phase at 4.3 V [17], [24]. Further, in our low-cobalt composition, the XRD diffractogram at 1.5 V discharge is best fitted by two *P6₃/mmc* phases with unequal crystallographic parameters. This is in contrast with a previous report of a single P6₃/*mmc* phase at low voltage during discharge for a composition with higher cobalt content [56]. We also observe an expansion of the c-axis parameter and a reduction of in-plane axes during discharge at low voltage. Detailed equivalent circuit modeling of Electrochemical Impedance Spectroscopy, combined with an analysis of the initial linearity of galvanostatic plots, reveals that the internal resistance of NMFCO, arising primarily from interfacial resistance at the cathode, does not increase with cycling as substantively as NMFO.

Combined with a discharge capacity of 195 mAhg$^{-1}$ in the 1.5 to 4.3 V range, a specific energy density exceeding 500 mWhg$^{-1}$, and relatively high state of health up to 200 cycles, we assert that NMFCO could be an excellent candidate for Na-based battery cathodes.

**Funding:**

This research did not receive any specific grant from funding agencies in the public, commercial, or not-for-profit sectors.

**Data Statement:**

Data used in this article is available directly from the authors. e-mail: pg@uwm.edu, rexhaus3@uwm.edu, and parson33@uwm.edu.

**Acknowledgements:**

We thank Dr. Steve Hardcastle for his support of our work at the Advanced Analysis Facility at the University of Wisconsin Milwaukee.

**Figure Captions:**

**Figure 1: Crystal structure of pristine uncycled NMFCO.** The subscripts on $Na_e$ and $Na_f$ refer to Na sites which are edge-sharing and face-sharing with the neighboring octahedra.

**Figure 2: X-ray Diffractograms (XRD) of NMFCO cathode at different charge or discharge voltages.** (a) XRD of NMFCO cathode material charged and discharged at different voltages (also see Table 1) during cycling in the voltage range 1.5–4.3 V (b) XRD of NMFCO at 1.5 V showing fits to *Cmcm* and *P6$_3$/mmc* phases.

**Figure 3. XRD diffractogram of uncycled and cycled NMFCO.** XRD of $Na_{0.67}Mn_{0.625}Fe_{0.25}Co_{0.125}O_2$ cathode surface after 200 cycles in the 1.5–4.0 V and 2.0–4.0 V ranges. Diffraction peaks for Aluminum 111 and 200 are showing in the 1.5-4.0V range, probably because a portion of the Aluminum current collector was (unintentionally) exposed during that XRD measurement.

**Figure 4. Discharge Capacity and Specific Energy of NMFO and NMFCO.** Shown here are measurements in voltage ranges (a, b) 1.5–4.3 V; (c, d) 1.5–4.0 V; and (e, f) 2.0–4.0 V.

**Figure 5. Galvanostatic cycling of NMFCO cathodes.** Cycling curves of NMFCO cathodes cycled in the voltage ranges 1.5-4.3 V, 1.5-4.0 V and 2.0-4.0 V.

**Figure 6. Initial slope of galvanostatic discharge curves as a function of cycle number.** The data was obtained from our galvanostatic charge-discharge measurements, including data shown in Figure 3 for different cycle numbers, for NMFO and NMFCO cathodes in the (a) 1.5-4.3 V, (b) 1.5-4.0 V and (c) 2.0-4.0 V ranges. Points depict best fit slopes in the 0-3.6 mAhg$^{-1}$ energy capacity range.

**Figure 7. Scanning Electron Micrography (SEM).** SEM images of $Na_{0.67}Mn_{0.625}Fe_{0.25}Co_{0.125}O_2$ (NMFCO) cathode surfaces as prepared, and after cycling. (a) Uncycled cathode; (b) after 200 cycles in the 1.5–4.0 V range; (c) after 200 cycles in the 1.5–4.0

V range. We conclude that formation of a Solid Electrolyte Interface (SEI) is observed in 'b' and 'c'.

**Figure 8. Electrochemical Impedance Spectroscopy results.** (a) Nyquist plots of NMFO (dashed line) and NMFCO (solid line). Data is shown for uncycled cells (black), and for cells after 10 cycles (red) and 200 cycles (blue) cycled in the 2–4 V range. All the spectra are obtained by varying frequency from 1000 kHz to 0.1 Hz. (b) Total impedance of uncycled and cycled (10 cycles, 200 cycles) cells of NMFO (red triangle) and NMFCO (green circle) as a function of negative log of frequency.

**Figure 9. Equivalent Circuit Fitting from Electrochemical Impedance Spectroscopy.** (a) Model A: Equivalent circuit diagram obtained after analysis of impedance spectra of cycled and uncycled NMFCO and NMFO half cells. (b) Model B: Equivalent circuit diagram describing a half cell with series bulk resistance and capacitance, diffusion of sodium ions inside the cathode, and charge transfer and double layer contributions from the cathode and interphase layer.